\documentclass{emulateapj}

\def\lsim{\lower.5ex\hbox{$\; \buildrel < \over \sim \;$}}
\def\gsim{\lower.5ex\hbox{$\; \buildrel > \over \sim \;$}}

\def\ch{\lower-0.55ex\hbox{--}\kern-0.55em{\lower0.15ex\hbox{$h$}}}
\def\lh{\lower-0.55ex\hbox{--}\kern-0.55em{\lower0.15ex\hbox{$\lambda$}}}

\shorttitle{Search for chaos in neutron stars: Is Cyg X-3 a black hole?}
\shortauthors{Karak, Dutta, Mukhopadhyay}

\begin{document}


\title{Search for chaos in neutron star systems:
Is Cyg X-3 a black hole?}


\author{Bidya Binay Karak\altaffilmark{1}, Jayanta Dutta\altaffilmark{2}, 
Banibrata Mukhopadhyay\altaffilmark{3}}
\affil{Astronomy and Astrophysics Program, Department of Physics,
Indian Institute of Science, Bangalore 560012, India}



\altaffiltext{1}{bidya$\_$karak@physics.iisc.ernet.in}
\altaffiltext{2}{dutta@physics.iisc.ernet.in}
\altaffiltext{3}{bm@physics.iisc.ernet.in}


\begin{abstract}
The accretion disk around a compact object
is a nonlinear general relativistic system involving magnetohydrodynamics.
Naturally the question arises whether such a system is
chaotic (deterministic) or stochastic (random) which might be related
to the associated transport properties whose origin is still not confirmed.
Earlier, the black hole system GRS~1915+105 was shown to be 
low dimensional chaos in certain temporal classes. 
However, so far such nonlinear phenomena have not been studied
fairly well for neutron stars which are unique for their
magnetosphere and kHz quasi-periodic oscillation (QPO). On the other hand, it 
was argued that the QPO is a result of nonlinear magnetohydrodynamic effects in 
accretion disks. If a neutron star exhibits chaotic signature, then 
what is the chaotic/correlation dimension? 
We analyze RXTE/PCA data of neutron stars Sco~X-1 and Cyg~X-2, along with the black hole 
Cyg~X-1 and the unknown source Cyg~X-3, and show that 
while Sco~X-1 and Cyg~X-2 are low dimensional chaotic systems, 
Cyg~X-1 and Cyg~X-3 are stochastic sources. 
Based on our analysis, we argue that Cyg~X-3 may be a black hole.
\end{abstract}


\keywords{stars: neutron ---  X-rays: binaries ---  X-rays: individual (Sco~X-1, Cyg~X-1,2,3)
--- accretion, accretion disks} 



\section{Introduction}


X-ray binary systems vary on timescales ranging from months to milli-seconds 
(see, e.g., \cite{chen, paul97, nowak, cui, gleis, axel0}). 
Detailed analysis of their temporal variability and fluctuation provides important 
insights into the geometry and physics of
emitting regions and the accretion process. However, the origin of variability is
still not clear. It could be due to varying external parameters, like the
infalling mass accretion rate. It could also be due to possible instabilities
in the inner regions of the accretion disk
where the flow is expected to be nonlinear and turbulent. 
Uttley et al. (2005) (see also Timmer et al. 2000 and Thiel et al. 2001) argued 
that the non-linear behavior of a system can be understood from the log-normal 
distribution of the fluxes and the $rms$-flux relation.
This implies that the temporal behavior of the system may be driven 
by underlying stochastic variations. By studying the underlying nonlinear behavior, 
important constraints can be obtained on these various possibilities. 

An elegant way of obtaining the constraint is to perform the nonlinear time 
series analysis of observed data
and to compute the correlation dimension $D_2$ in a non-subjective manner.
This technique has already been used to diverse situations 
(\citep{grass0,grass,schr,aber,skb,bmc,misra,hari}, and references therein).
By obtaining $D_2$ as a function of the embedding dimension $M$, one can infer the origin of
the variability. For example, $D_2\approx M$ for all $M$ corresponds to the system having
stochastic fluctuation which favors the idea that X-ray variations are driven by variations 
of some external parameters. 
On the other hand, a saturated $D_2$ to a finite (low) value, beyond a certain M, implies 
a deterministic chaos which argues in favor of inner disk instability. 
However, to implement the algorithm successfully, 
the system in question should provide enough data. 

The technique was used earlier to understand the nonlinear nature of 
a black hole system Cyg~X-1 \citep{unno} and an Active Galactic Nucleus (AGN)
Ark~564 \citep{agn}, but due to insufficient data points the analyses were hampered
and no concrete conclusions were made about $D_2$. Later on, another
black hole system GRS~1915+105 was analyzed \cite{bmc,misra,hari} which 
was shown to display low dimensional chaos in certain
temporal classes, while stochastic in other classes.

However, so far none of the neutron star systems have been analyzed 
in detail in order to understand the origin of nonlinearity.
Decades back, Voges et al. (1987) attempted to understand the chaotic nature of Her~X-1,
but the analysis was hampered by low signal to noise ratio
\cite{noris}.
Since then the investigation
of chaotic signature in neutron stars remains unattended.
Can a neutron star system not be deterministic?  
Indeed several features of X-ray binary systems 
consisting of a neutron star, such as their 
magnetosphere and kHz Quasi-Periodic Oscillation (QPO) and its
possible relation to the spin frequency of the neutron star, favor the
idea that they exhibit nonlinear resonance (e.g. \cite{blaes,bm}).
While the QPO itself is a mysterious feature whose origin is still unclear,
its possible link to the spin frequency of the neutron star\footnote{However, 
some authors (Mendez \& Belloni 2007) suggested that the kHz QPOs may
not be related to the spin.} indicates 
the origin of QPO to be from nonlinear phenomena.
Several LMXBs having a neutron star exhibit twin kHz 
QPOs \cite{mendez, vander}. For some of them, e.g.
4U~1636-53 \cite{jonker}, KS~1731-260 \cite{smith}, 4U~1702-429 
\cite{markwardt}, 4U~1728-34 \cite{vans}, the spin frequency 
of the neutron star has been predicted from observed data. 
However, for the source Sco~X-1, which
exhibits noticeable time variability \cite{menvan}, 
while we observe twin kHz QPOs, we do not know the spin frequency yet
(but see \cite{bm}). For another neutron star Cyg~X-2, we do observe
kHz QPOs \cite{wijnands} as well.
Several black holes also exhibit QPOs, e.g. GRS~1915+105
\cite{belloniqpo, mcclintock}, Cyg~X-1 \cite{angelini}.

In the present paper, we first aim at analyzing the time series of
two neutron star sources Sco~X-1 and Cyg~X-2 to understand if a neutron star is
a deterministic nonlinear (chaotic) system. Then we try to manifest
the knowledge of nonlinear (chaotic/random) property of compact sources 
to distinguish a black hole from a neutron star.
Subsequently, knowing their difference based on the said 
property, we try to identify the nature of a unknown source
(whether it is a black hole or a neutron star).
While the nature of some sources, as mentioned above, has
already been predicted based on alternate method, for some others,
e.g. Cyg~X-3, SS433, it has not yet been confirmed.

For the present purpose, we therefore concentrate on three additional sources 
Cyg~X-1, Cyg~X-2 and Cyg~X-3. 
While Cyg~X-1 has been predicted to be a black hole and Cyg~X-2 be a neutron star,
the nature of Cyg~X-3 is not confirmed yet. Some authors 
\cite{ergma98,schmutz96,szostek08} argued for a black hole nature of Cyg X-3, 
on the basis of its jet, the time variations in the infrared emission lines, 
the BeppoSAX X-ray spectra and so on.
However, earlier it was argued for a neutron star \cite{chadwick85}
by measuring its $1000$ GeV $\gamma$-rays which suggests a pulsar period of $12.5908\pm 0.0003$ 
ms. By analyzing the time series and computing the correlation dimension $D_2$, 
here we aim at pinpointing the nature of Cyg~X-3: whether a black hole or 
a neutron star. 

In the next section, we briefly outline the procedure to be followed in understanding 
the nonlinear nature of a compact object from observed data and to implement it to analyze 
the neutron star source Sco~X-1. 
In \S3, we then describe nonlinear behaviors
of Cyg~X-1, Cyg~X-2 and Cyg~X-3. Subsequently, in \S4, we compare all the results
and argue for a black hole nature of Cyg~X-3. Finally, we summarize in \S5. 

\section{Procedure and nonlinear nature of Sco~X-1}

The method to obtain $D_2$ is already established (see, e.g., \citep{grass,bmc,hari}).
Therefore, here we discuss it briefly. We consider PCA data of the RXTE satellite 
(see Table 1 for the observations IDs) 
from the archive for our analysis. We process the data using the FTOOLS software. 
We extract a few continuous data streams of $2500-3500$ sec long.
The time resolution used to generate lightcurves 
is $\sim 0.1-1$ sec. This is the range of optimum resolution, at least for 
the sources we consider, to minimize noise without losing physical 
information of the sources. A finer time resolution
would be Poisson noise dominated and a larger binning might give too few data points 
to derive physical parameters from it (see Misra et al. 2004, 2006, for details). 

Then we calculate the correlation dimension according to the Grassberger $\&$ Procaccia (1983a,b) 
algorithm. From the time series $s(t_i)$ (i = 1,2,...,N), we construct an M dimensional space 
(called embedding space ), in which any vector has the following form:
\begin{equation}
x(t_{i})=[s(t_{i}),s(t_{i}+\tau), ........., s(t_{i}+(M-1)\tau)],
\end{equation}
where $\tau$ is the time delay chosen in such a way that each component of the
vector $x(t_i)$ is 
independent of each other. For a particular choice of embedding dimension $M$, we 
compute the correlation function:

\begin{equation}
C_M(r)=\frac{1}{N(N_c-1)}\sum_{i=1}^{N} \sum_{j=1,j\neq i}^{N_c}\Theta(r-|x_{i}-x_{j}|),
\end{equation}
which is basically the average number of points within a hypersphere of diameter $r$,
where $\Theta$ is a Heaviside step function, $N$ the total number of points 
and $N_c$ the number of centers. If the system has a strange attractor, then 
one can show that for a small value of $r$ 
\begin{equation}
{D_{2}(M)}= \frac{d~ {\rm log}~C_M(r)}{d~ {\rm log}~r}.
\end{equation}

\begin{figure}
\epsscale{.80}
\plotone{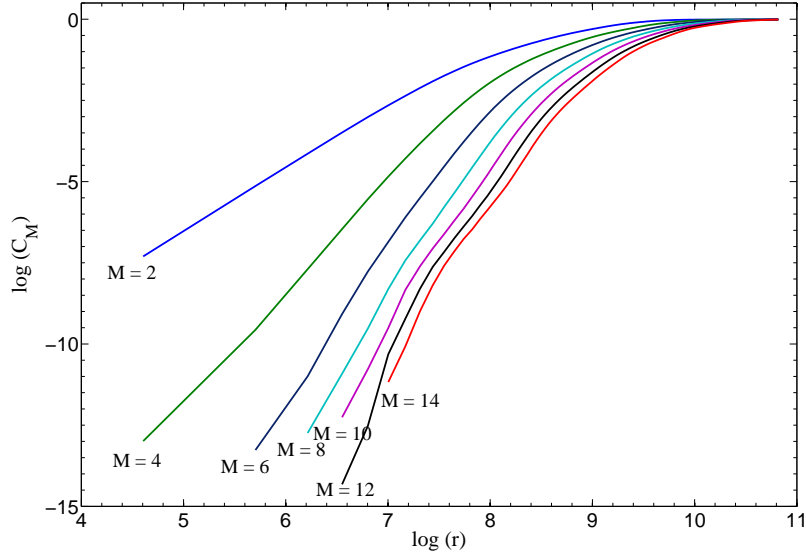}
\caption{Variation of log ($C_M$) as a function of log($~r$) for different 
embedding dimensions. The linear scaling range is used to 
calculate the correlation dimension.
\label{fig6}}
\end{figure}

In this numerical calculation, we divide the whole phase space into $M$ cubes of length
$r$ around a point and we count the average number of data points in these cubes to calculate $C_M(r)$. The edge
effects, which come due to the finite number of data points, have been avoided by calculating
$C_M(r)$  in the range $ r_{min}<r<r_{max} $, where $r_{min}$ is the value of $r$ for 
$C_M(r)$ just greater than one and $r_{max}$ can be found by restricting the $M$
cubes to be within the embedding space. In Fig. \ref{fig6}, we show the variation of log$(C_M(r))$ with 
log$(r)$ for different values of $M$ for Sco X-1 data.

$D_{2}(M)$ can be calculated from the linear part of the ${\rm log}(C_M(r))$ vs. ${\rm log}(r)$ curve
and its value depends on the value of $M$. For a stochastic system, 
$D_{2} \approx M$ for all $M$. On the other hand, for a chaotic or deterministic system, 
initially $D_{2}(M)$ increases linearly with the increase
of $M$, then it reaches a certain value and saturates. This saturated value of $D_{2}$
is taken to be the correlation dimension of the system which is a non-integer.
The standard deviation gives the error in $D_{2}$.

\begin{figure}
\epsscale{.80}
\plotone{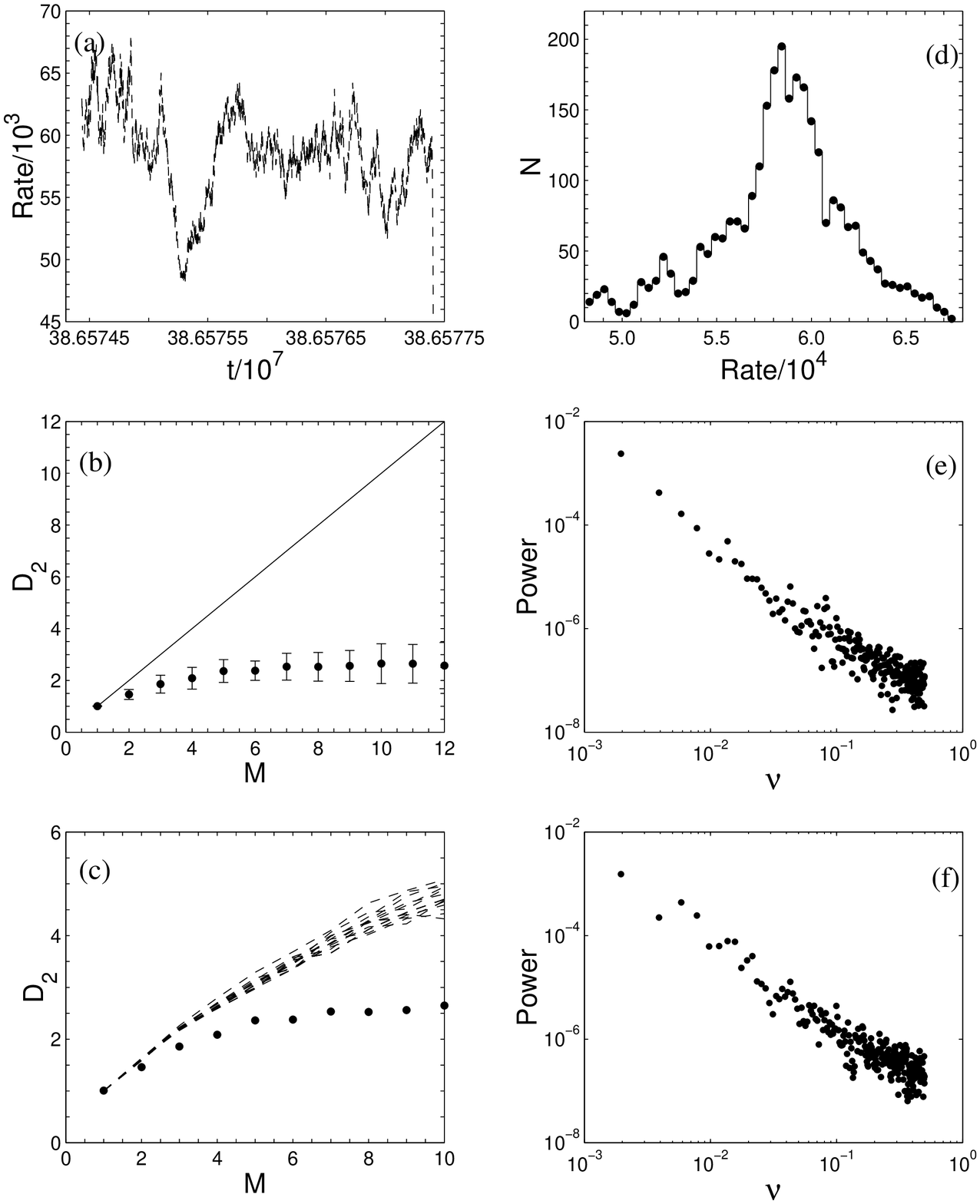}
\caption{
 Sco~X-1: (a) Variation of count rate as a function of time in units of $10^7$ sec
(lightcurve), without subtracting the initial observation time.
(b) Variation of correlation dimension, along with error bars,
as a function of embedding dimension for
original data. The solid line
along the diagonal of the figure indicates an ideal stochastic curve.
(c) Variation of correlation dimension as a function of embedding dimension
for original (points) and corresponding surrogate (dashed lines) data.
(d) Variation of number of count
rate as a function of count rate itself in units of $10^4$ sec$^{-1}$ (Distribution) for original
(solid line) and surrogate (points) data. Power-spectra for (e) original and (f)
surrogate data.
\label{fig1}}
\end{figure}

We first concentrate upon the neutron star source Sco~X-1. 
In Figs. \ref{fig1}a,b we show respectively 
the lightcurve and the variation of 
$D_2$ as a function of $M$ and find that
$D_2$ saturates to a value $ 2.6~ (\pm~ 0.8)$. As this is a non-integer, 
the system might be chaotic. On the other hand, we know that the 
Lorenz attractor is an example of an ideal chaos with $D_2=2.05$. 
Therefore, Sco~X-1 may be like a Lorenz system. But due to noise its $D_2$ 
seems appearing slightly higher \cite{bmc, misra} than the actual value. However, one should be cautious
about the fact that Sco~X-1 is a bright source (much brighter than other sources 
considered later). Hence, the dead time effect on the detector might affect
the actual value of saturated $D_2$ and the computed value might be slightly 
different than the actual one. However, this can not rule out the signature 
of chaos in Sco~X-1, particularly because the corresponding count rates are 
confined in the same order of magnitude and hence the dead time effect, if any,
is expected to affect all the count rates in a similar way.

However, a saturated $D_2$ is necessary but not a sufficient evidence for chaos.
Existence of color noise (for which the power spectrum
$P(\nu) \propto \nu^{-\alpha}$, where the power spectral indices $\alpha$ = 0, 1 
and 2 correspond to ``white", ``pink" and ``red" noise respectively) into a stochastic 
system might lead to a saturated $D_2$ of 
low value as well (e.g. Osborne \& Provenzale 1989; Theiler et al 1992; Misra et al. 2006; 
Harikrishnan et al. 2006). 
Therefore, it is customary to analyze data by alternate approach(s) to distinguish it
from a pure noisy time series \citep{kuz}. One of the 
techniques is the surrogate data analysis (e.g. \citep{schreiber96}), 
which has been described earlier in detail
and implemented for a black hole \citep{misra, hari}. 
In brief, surrogate data is random data generated by taking the original signal and reprocessing it
so that data has the same/similar Fourier power spectrum and autocorrelation 
along with the same distribution, mean and variance
as of the original data, but has lost all deterministic 
characters. Then the same analysis is carried out to the original data and 
the surrogate data to identify any distinguishable feature(s) between them. 
The scheme proposed by Schreiber \& Schmitz (1996), known as Iterative Amplitude-Adjusted Fourier 
Transform (IAAFT), is more consistent to generate surrogate data.


Figures \ref{fig1}c-f compare results for the original data with the surrogate data.
It is clear that while distributions and power spectra are same/similar for
both the data sets, $D_2$ is much higher for the surrogate data which suggests
existence of low dimensional chaos in Sco~X-1 with $D_2\sim 2.6$. 
This confirms, for the first time to best
of our knowledge, a neutron star source to display chaotic behavior. 
As the existence of chaos is a 
plausible signature of instability in the inner region of accretion flows which
is nonlinear and turbulent, as mentioned in \S 1, the corresponding QPO, which 
is presumably an inner disk phenomenon as well, is expected
to be governed by nonlinear resonance mechanisms (e.g. Mukhopadhyay 2009).

In Table 1, we enlist the average counts $<S>$, its root mean square ($rms$) variation
$\sqrt{<S^2>-<S>^2}/<S>$, the expected Poisson noise $<PN>~\equiv \sqrt{<S>}$,
and the ratio of the expected Poisson noise to the $rms$ value for all sources. It clearly
shows a strong correlation between the inferred behavior of the systems and the
ratio of the expected Poisson noise to the $rms$ value.

\section{Nonlinearity of Cyg~X-1,2,3}

\begin{figure}
\epsscale{.80}
\plotone{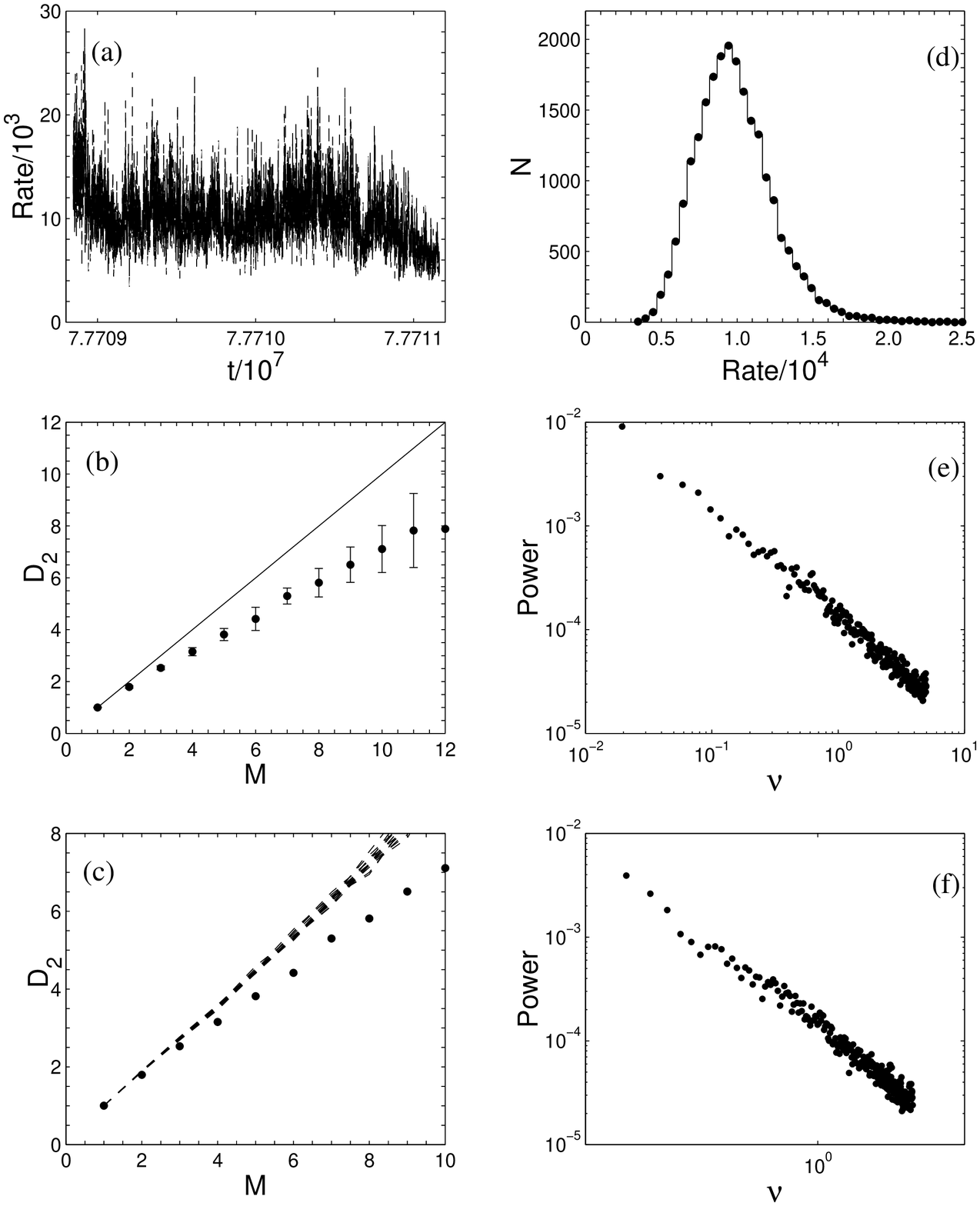}
\caption{Cyg~X-1: Same as Fig. \ref{fig1}.
\label{fig2}}
\end{figure}

We now look into three additional compact sources: Cyg~X-1 (black hole), Cyg~X-2 (neutron star) and Cyg~X-3
(nature is not confirmed yet), and apply the same analysis as in the case of Sco~X-1. 
Figure \ref{fig2}b shows that
$D_2$ for Cyg~X-1 seems not to saturate and appears very high. However, there is no surprise in it
because its variability is similar to the temporal class $\chi$ of the black hole GRS~1915+105
which was shown to be Poisson noise dominated and stochastic in nature \cite{bmc}.
Indeed, earlier analysis of Cyg~X-1 data,
while it could not conclusively quantify the underlying chaotic
behavior due to insufficient data, revealed very
high dimensional chaos.
Moreover a large $<PN>$ (as well $<PN>/rms$) for Cyg~X-1, compared to that for Sco~X-1 
given in Table 1, reveals the system to be noise dominated. This ensures Cyg~X-1 to 
be different from Sco~X-1. However, the variation of $D_2$ as a function of $M$ for 
the original data does not deviate noticeably from that of corresponding surrogate data, as 
shown in Fig. \ref{fig2}c, which argues that Cyg~X-1 is not a chaotic system. 

\begin{figure}
\epsscale{1.0}
\plotone{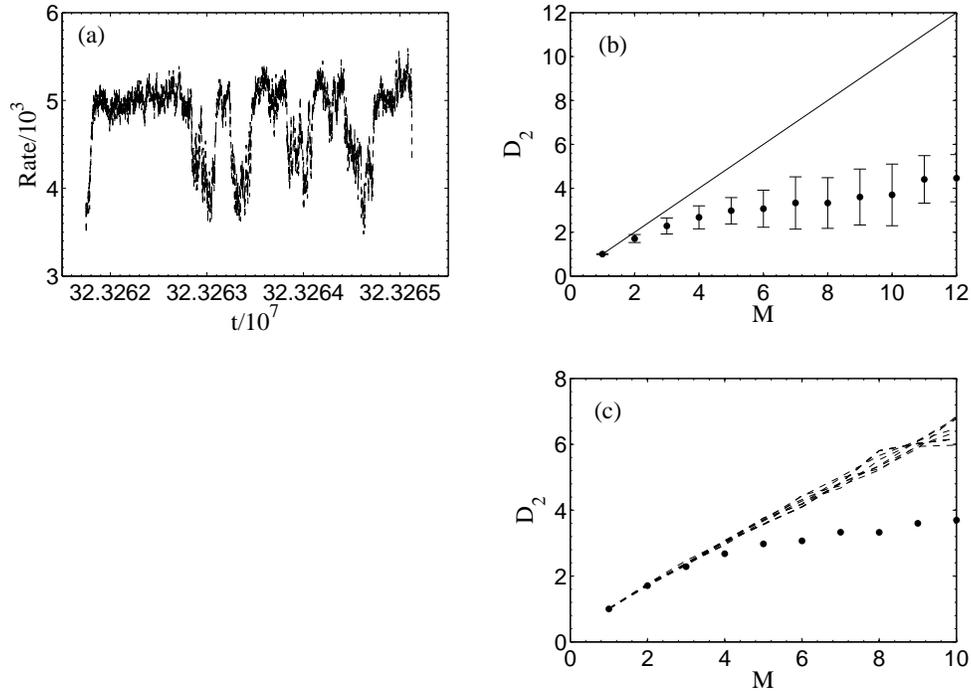}
\caption{
Cyg~X-2: (a) Variation of count rate as a function of time in units of $10^7$ sec
(lightcurve).
(b) Variation of correlation dimension, along with error bars,
as a function of embedding dimension for
original data. The solid line
along the diagonal of the figure indicates an ideal stochastic curve.
(c) Variation of correlation dimension as a function of embedding dimension
for original (points) and corresponding surrogate (dashed lines) data.
\label{fig3}}
\end{figure}

Figures \ref{fig3}b,c show that $D_2$ for Cyg~X-2 saturates to a low value $\sim 4$, 
which is significantly
different than that of corresponding surrogate data. The power spectra and distributions, 
on the other hand, for original and surrogate data are same/similar (as shown in Sco~X-1,
not repeated further). The saturated $D_2$ for Cyg~X-2 
is almost double than that of Lorenz system, possibly due to high Poisson noise to $rms$ 
ratio (see Table-1). This suggests the corresponding system to be a low dimensional 
chaos. 

\begin{figure}
\epsscale{1.0}
\plotone{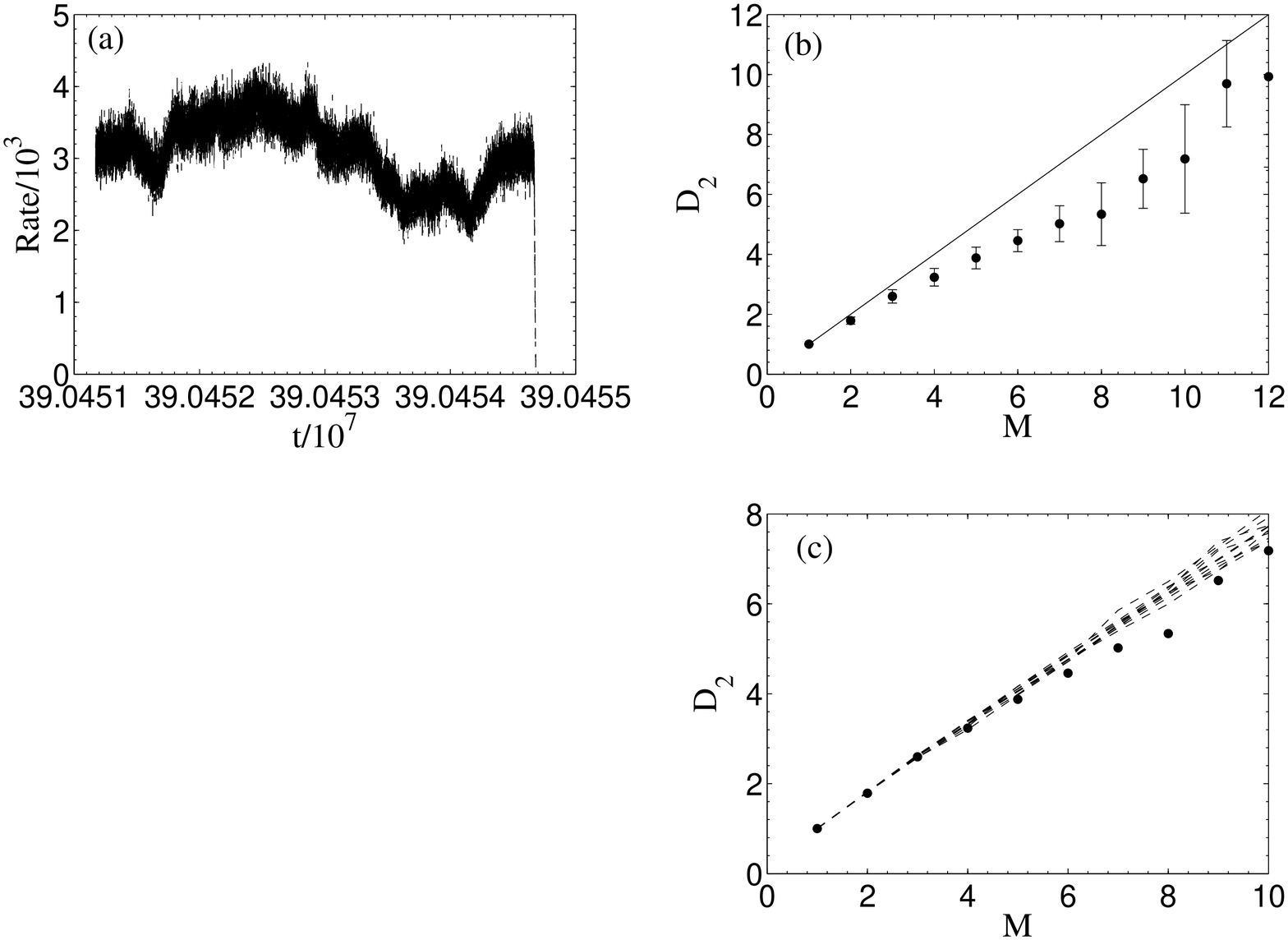}
\caption{
Cyg~X-3: Same as Fig. \ref{fig3}.
\label{fig4}}
\end{figure}

From Figs. \ref{fig4}b,c we see that for 
Cyg~X-3 the variations of $D_2$ as a function of $M$ for original and surrogates data are similar 
to that of Cyg~X-1. This confirms that the behavior of the unknown source Cyg~X-3 is 
similar to that of the black hole source Cyg~X-1 (see, however, \cite{axel}). 
Note from Table 1 that Cyg~X-1, X-2, X-3 
are significantly noise dominated compared to Sco~X-1. Although noise could not
suppress the chaotic signature in the neutron star Cyg~X-2, its saturated $D_2$ is 
higher than that of Sco~X-1. On the other hand, even though the Poisson noise to $rms$ 
ratio in Cyg~X-1 is lower than that in Cyg~X-2 (but Poisson noise itself
is higher in Cyg~X-1), its $D_2$ never saturates, which
confirms the source to be non-chaotic; the apparent stochastic signature is not 
due to Poisson noise present into the system.

\section{Comparison between Cyg~X-1, Cyg~X-2 and Cyg~X-3}

\begin{figure}
\epsscale{.80}
\plotone{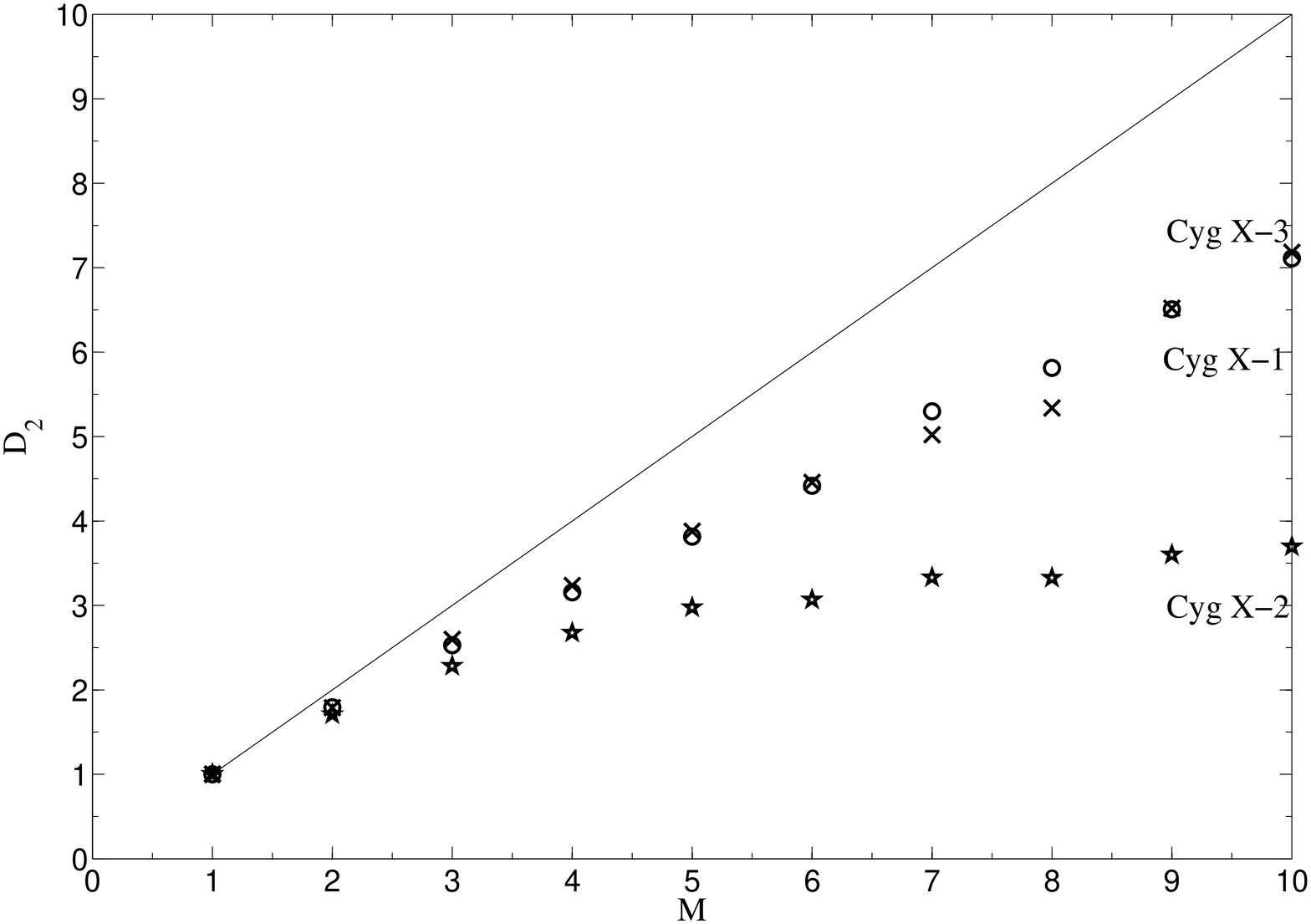}
\caption{
Comparison of the variation of correlation dimension as
a function of embedding dimension between Cyg~X-1 (open circle), Cyg~X-2 (star), Cyg~X-3 (cross).
\label{fig5}}
\end{figure}

Finally, we compare the variations of $D_2$ for all three cases
of Cygnus in Fig. \ref{fig5}. Remarkably we find that $D_2$ values for Cyg~X-1 and Cyg~X-3 
practically overlap, appearing much larger compared to that for Cyg~X-2
which is shown to be a low dimensional chaotic source. 
On the other hand, Cyg~X-2 is a confirmed neutron star and Cyg~X-1 a black hole.
Therefore, Cyg~X-3 may be a black hole. 

\section{Summary}
The source Cyg~X-3, whose nature is not confirmed
yet, seems to be a black hole based on the analysis of its nonlinear behavior. 
On the other hand, we have shown, for the first 
time to best of our knowledge, that 
neutron star systems could be chaotic in nature. The signature of deterministic chaos,
which argues in favor of inner disk instability, into
an accreting system has implications in understanding its transport properties 
particularly in Keplerian accretion disk \cite{win}.
Note that in Keplerian accretion disks
transport is necessarily due to turbulence in
absence of significant molecular viscosity. The signature of chaos confirms
instability and then plausible turbulence.
On the other hand, for a rotating neutron star having a magnetosphere, signature 
of chaos suggests their QPOs to be nonlinear resonance phenomena \cite{bm}. 
The absence of chaos and related/plausible signature of instability in Cyg X-1 and Cyg X-3 
suggests the underlying accretion disk to be sub-Keplerian \cite{ny95,chak96} in nature 
which is dominated significantly by gravitational force.



\acknowledgments

This work is partly supported by a project (Grant No. SR/S2/HEP12/2007)
funded by Department of Science and Technology (DST), India. Also the financial support to
one of the authors (JD) has been acknowledged. The authors would like to thank 
Arnab Rai Choudhuri of IISc and the anonymous referee for carefully reading the 
manuscript, constructive comments and suggestions.

\clearpage

\begin{table*}[htbp]
\caption{Observed data}
\begin{tabular}{cccccccccccccccccccccc}\\
\hline
\hline
Source      & Obs. I. D.      & $<S>$  &
$rms$          & $<PN>$    &
$<PN>/rms$  & Behavior  \\
\hline
\hline
Sco~X-1 & 91012-01-02-00  & $58226$ & $0.074$ & $0.004$ & $0.054$ & C \\
\hline
Cyg~X-1 & 10512-01-09-01 &     $10176$ & $0.261$ & $0.031$ & $0.119$ & NC/S \\
\hline
Cyg~X-2 & 10063-10-01-00 & $4779$ & $0.075$ & $0.014$ & $0.191$ & C \\
\hline
Cyg~X-3 & 40061-01-07-00 &   $3075$ & $0.125$ & $0.057$ & $0.455$ & NC/S \\ 
\hline
\hline
\end{tabular}\\ \\
{Columns:- 1: Name of the source,
2: RXTE Observational identification number from
which the data has been extracted. 
3: The average count in the lightcurve $<S>$ 4: The root mean square variation
in the lightcurve, $rms$.
5: The expected Poisson noise variation, $<PN> \equiv \sqrt{<S>}$.
6: The ratio of the expected Poisson noise to the actual root mean
square variation 7: The behavior of the system (C: chaotic behavior; 
S: stochastic behavior; NC: nonchaotic behavior)}
\end{table*}


\clearpage

\clearpage






\begin{thebibliography}{}

\bibitem[Aberbandel 1996]{aber} Aberbandel, H. D. L. 1996, {\it Analysis of
Observed Chaotic data} (Springer: New York)

\bibitem[Angelini et al. 1994]{angelini}Angelini, L., White, N. E., Stella, L. 1994, in New Horizon of X-Ray Astronomy, ed. F. Makino, \& T. Ohashi (Tokyo: Universial Academy Press), 429

\bibitem[Axelsson 2008]{axel0} Axelsson, M. 2008, AIPC, 1054, 135

\bibitem[Axelsson et al. 2008]{axel} Axelsson, M., Larsson, S. \& Hjalmarsdotter, L. 2008, MNRAS, 394, 1544

\bibitem[Belloni et al. 2001]{belloniqpo} Belloni, T., M\'endez, M., S\'anchez-Fern\'andez, C. 2001, ApJ, 372, 551

\bibitem[Blaes et al. 2007]{blaes} Blaes, O. M., Srámkov\'a, E., Abramowicz, M. A., 
Kluźniak, W., Torkelsson, U. 2007, ApJ, 665, 642

\bibitem[Chadwick et al. 1985]{chadwick85} Chadwick, P. M., Dipper, N. A.,  
Dowthwaite, J. C., Gibson, A. I. \& Harrison, A. B. 1985, Nature, 318, 642

\bibitem[Chakrabarti 1996]{chak96} Chakrabarti, S. K. 1996, ApJ, 464, 664

\bibitem[Chen et al. 1997]{chen} Chen, X., Swank, J. H. \& Taam, R. E. 1997, ApJ, 477, L41.

\bibitem[Cui 1999]{cui} Cui, W. 1999, ApJ, 524, L59

\bibitem[Ergma \& Yungelson 1998]{ergma98} Ergma, E. \& Yungelson, L. R. 1998, A\&A, 333,
151

\bibitem[Gleissner et al. 2004]{gleis} Gleissner, T., Wilms, J., Pottschmidt, K., Uttley, P., 
Nowak, M. A., \& Staubert, R. 2004, A\&A, 414, 1091

\bibitem[Gliozzi et al. 2002]{agn} Gliozzi, M., Brinkmann, W., R\"ath, C., Papadakis, 
I. E., Negoro, H. \& Scheingraber, H. 2002, A\&A, 391, 875


\bibitem[Grassberger \& Procaccia 1983a]{grass0} 
Grassberger, P. \& Procaccia, I. 1983, Physica D, 9, 189

\bibitem[Grassberger \& Procaccia 1983b]{grass} Grassberger, P. \& Procaccia, I 1983,
Phys. Rev. Lett., 50, 346

\bibitem[Harikrishnan et al. 2006]{hari} Harikrishnan, K. P., Misra, R., Ambika, G.
\& Kembhavi, A. K. 2006, Physica D, 215, 137


\bibitem[Jonker et al. 2002]{jonker} Jonker, P. G., Mendez, M., \& van der Klis, M. 2002, MNRAS, 336, L1

\bibitem[Kugiumtzis 1999]{kuz} Kugiumtzis, D. 1999, Phys. Rev. E, 60, 2808


\bibitem[Markwardt et al. 1999]{markwardt} Markwardt, Craig B., Strohmayer, Tod E. \& Swank, Jean H, 1999, ApJ 512, L125

\bibitem[McClintock \& Remillard 2006]{mcclintock} McClintock, J. E., \& Remillard, R. A. 2006, in Compact Stellar X-Ray Sources, ed. W. H. G. Lewin \& M. van der Klis, (Cambridge: Cambridge Univ. Press)

\bibitem[Mendez \& Belloni 2007]{menbel07} Mendez, M., \& Belloni, T. 2007, MNRAS, 381, 790

\bibitem[Mendez et al. 1998]{mendez}Mendez, M., van der Klis, M., Wijnands, R., Ford, E. C., van Paradijis, J., \& Vaughan, B. A. 1998, ApJ, 505, L23

\bibitem[Mendez \& van der Klis 2000]{menvan}Mendez, M. \& van der Klis 2000, MNRAS 318, 938


\bibitem[Misra et al. 2006]{misra} Misra, R., Harikrishnan, K. P., Ambika, G. 
\& Kembhavi, A. K. 2006, ApJ, 643, 1114

\bibitem[Misra et al. 2004]{bmc} Misra, R., Harikrishnan, K. P., Mukhopadhyay, B., 
Ambika, G. \& Kembhavi, A. K. 2004, ApJ, 609, 313

\bibitem[Mukhopadhyay 2009]{bm} Mukhopadhyay, B. 2009, ApJ, 694, 387

\bibitem[Narayan \& Yi 1995]{ny95} Narayan, R. \& Yi, I. 1995, ApJ, 452, 710

\bibitem[Norris \& Matilsky 1989]{noris} Norris, J. P. \& Matilsky, T. A. 1989, ApJ, 346, 912 

\bibitem[Nowak et al. 1999]{nowak} Nowak, M. A., Vaughan, B. A., Wilms, J., Dove, J. B. \& Begelman, M. C. 1999, ApJ, 510, 874

\bibitem[Osborne \& Provenzale 1989]{osb} Osborne, A. R. \& Provenzale, A. 1989, Phy. D, 35, 357



\bibitem[Paul et al. 1997]{paul97} Paul, B., Agrawal, P. C., Rao, A. R., Vahia, M. N., Yadav, J. S., Marar, T. M. K., Seetha, S., Kasturirangan, K. 1997, A\&A 320 L37

\bibitem[Schmutz et al. 1996]{schmutz96} Schmutz, W., Geballe, T. R. \& Schild, H.
1996, A\&A, 311, 25

\bibitem[Schreiber 1999]{schr}
Schreiber, T. 1999, Phys. Rep., 308, 1

\bibitem[Schreiber \& Schmitz 1996]{schreiber96} Schreiber, T. \& Schmitz, A. 1996,
Phys. Rev. Lett., 77, 635


\bibitem[Serre et al. 1996]{skb} Serre, T., Kollath, Z. \& Buchler, J. R. 1996, 
A\&A, 311, 833


\bibitem[Smith et al. 1997]{smith} Smith, D. A., Morgan, E. H., \& Bradt, H. 1997, ApJ, 479, 137

\bibitem[Szostek \& Zdziarski 2008]{szostek08} Szostek, A. \& Zdziarski, A. A. 2008,
MNRAS, 386, 593

\bibitem[Theiler et al. 1992]{theiler92} Theiler, J., Eubank, S., Longtin, A., Galdrikian, B., Doyne, F. J. 1992, Physica D, 58, 77

\bibitem[Unno et al. 1990]{unno} Unno, W., et al. 1990, PASJ, 42, 269

\bibitem[Timmer al. 2000]{timmer} Timmer, J, Schwarz, U \& Voss, H. U et al. 2000, Phys. Rev. E, 61, 1342

\bibitem[Thiel et al., 2001]{thiel} Thiel, M., Romano, M \& Schwarz, U et al. 2001, A\&A suppl. 276, 187

\bibitem[Uttley et al. 2005]{uttley} Uttley, P., McHardy, I. M., Vaughan, S. 2005, MNRAS, 359, 345
 
\bibitem[van der Klis 2006]{vander} van der Klis, M. 2006, AdSpR, 38, 2675

\bibitem[van Straaten et al. 2002]{vans} van Straaten, S., van der Klis, M., di Salvo, T., \& Belloni, T. 2002, ApJ, 568,
912

\bibitem[Voges et al. 1987]{her} Voges, W., Atmanspacher, H., \& Scheingraber, H. 1987, ApJ, 320, 794

\bibitem[Winters et al. 2003]{win} Winters, W. F., Balbus, S. A., \& Hawley, J. F. 2003,
MNRAS, 340, 519


\bibitem[Wijnands et al. 1998]{wijnands} Wijnands, R., Homan, J., van der Klis, M., Kuulkers, E., van Paradijs, J., 
Lewin, W. H. G., Lamb, F. K., Psaltis, D. \& Vaughan, B. 1998, ApJ, 493, L87

\end{thebibliography}
\end{document}